\documentclass[prd,onecolumn,superscriptaddress,amsfonts,amssymb,amsmath,showpacs]{revtex4-2}
\usepackage{bm}
\usepackage{amsfonts}
\usepackage{latexsym}
\usepackage[latin1]{inputenc}
\usepackage{graphicx}
\usepackage{amsmath}
\usepackage{palatino}
\usepackage{mathpazo}
\usepackage{textcomp}
\usepackage{float}
\usepackage{booktabs}
\usepackage{dcolumn}
\usepackage{booktabs}
\usepackage{multirow}
\usepackage{hyperref}
\hypersetup{
    colorlinks=true,
    linkcolor=purple,
    filecolor=magenta,      
    citecolor=blue
}
\usepackage{amsmath}
\usepackage{xcolor}
\usepackage{orcidlink}
\usepackage[caption=false]{subfig}
\usepackage{commath}
\makeatletter
\long\def\dddddot#1{%
  {\mathop {#1}\limits ^{\vbox to-1.4\ex@ {\kern -\tw@ \ex@ \hbox {\normalfont .....}\vss }}}%
}
\long\def\multidots#1#2{%
  \count@=0
  {{\mathop {#2}\limits ^{\vbox to-1.4\ex@ {\kern -\tw@ \ex@ \hbox {\normalfont %
  \loop%
  \ifnum#1>\count@%
  .%
  \advance\count@ by1%
  \repeat%
  }\vss }}}}%
}
\makeatother

\begin{document}
%\input epsf.tex
%%%%%%%%%%%%
%%%%%%%%%%%

\title{\bf Casimir wormhole with GUP correction in extended symmetric teleparallel gravity }

\author{Abhilipsa Sahoo\orcidlink{0009-0005-4381-3794}}
\email{abhilipsasahoo5862@gmail.com}
\affiliation{Department of Physics, Indira Gandhi Institute of Technology, Sarang, Dhenkanal, Odisha-759146, India.}

\author{S.K. Tripathy\orcidlink{0000-0001-5154-2297}}
\email{tripathy\_sunil@rediffmail.com}
\affiliation{Department of Physics, Indira Gandhi Institute of Technology, Sarang, Dhenkanal, Odisha-759146, India.} 
 
\author{B. Mishra \orcidlink{0000-0001-5527-3565}}
\email{bivu@hyderabad.bits-pilani.ac.in}
\affiliation{Department of Mathematics, Birla Institute of Technology and Science-Pilani, Hyderabad Campus, Hyderabad-500078, India.}

\author{Saibal Ray\orcidlink{0000-0002-5909-0544}}
\email{saibal.ray@gla.ac.in}
\affiliation{Centre for Cosmology, Astrophysics and Space Science (CCASS), GLA University, Mathura 281406, Uttar Pradesh, India.}

\begin{abstract}
{\bf Abstract:} Quantum mechanical concept such as the Casimir effect is explored to model traversable wormholes in an extended teleparallel gravity theory. The minimal length concept leading to the generalized uncertainty principle (GUP) is used to obtain the Casimir energy density. The effect of the GUP correction in the geometrical and physical properties of traversable Casimir wormholes are investigated. It is noted that the GUP correction has a substantial effect on the wormhole geometry and it modifies the energy condition. From a detailed calculation of the exotic matter content of the GUP corrected Casimir wormhole, it is shown that, a minimal amount of exotic matter is sufficient to support the stability of the wormhole.
\end{abstract}

\maketitle
\textbf{Keywords}:  Casimir Wormhole, Traversable Wormhole, $f(Q,T)$ Gravity, Energy Conditions.

\section{Introduction}

Wormholes can be defined as the hypothetical tunnel connecting two asymptotic regions of the same space time. To be very specific, wormhole is any compact region of space time with topological boundary, however its interior is topologically non-trivial \cite{Visser95}. These hypothetical bridges, known as the Einstein-Rosen bridges \cite{Einstein35a, Einstein1935b} were obtained as solutions of General Relativity (GR) \cite{Morris88}.  However, due to the recent observational signatures of black holes ~\citep{Oldham2016,Abbott2016,Bouman2016,EHT2019}, the related
concept of wormhole has come into the field of attraction to the researchers. As a result, specifically construction of
traversable wormholes are nowadays affluently available in GR as well as in modified gravity theories~\citep{Bejarano2017,Moraes2017,Cremona2019,Garattini2019,Tripathy2021,Mishra2021b,Chakraborty2021,Mishra2022,Sengupta2022,Mustafa2022,Mustafa2023,Chakraborty2023}. 

The traversable wormholes contain exotic matter with negative energy density and it is quite obvious that they violate the positivity condition of the sum of the energy density ($\rho$) and pressure ($p$). This condition is known as the Null Energy Condition $\rho+p\geq 0$ (NEC). The possibility of physical travel through the wormhole tunnel depends on the opening of its mouth which requires the matter content to be exotic. In fact, in his work, Visser \cite{Visser95} attempted to avoid falling of the traversable path within the exotic matter area. Since, within the classical Physics, it is not possible to create matters with negative energy density, in principles, traversable wormholes should not exist classically. Though there are many attempts made for the existence of traversable wormholes, but there is no detection or even trace of traversable wormholes. Recently it has been indicated that Casimir energy with negative energy density can be a potential physical source for the existence of traversable wormhole \cite{Butcher14,Garattini19}. The negative energy density due to the Casimir effect is a manifestation of the quantum fluctuation of the vacuum of the electromagnetic field between two plane parallel, uncharged conducting plates \cite{Casimir48}. Subsequently, the modelling of traversable wormhole exploring the Casimir effect has been shown in extended theory of gravity \cite{Tripathy2021}. There are some more studies available in literature \cite{Javed20,Jusufi20, Zubair2023, Rani2023, Samart2022, Shweta2023,Mishra2023} on Casimir wormhole by employing different modified theories of gravity. It is worthy to mention here that the Casimir energy represents the artificial yet laboratory source of exotic matter and it has strong dependence on the geometry of boundaries. However, the result of this Casimir effect is in principle but not in practice.\\  

The curvature and teleparallel representation are two equivalent geometric representation of General Relativity, the other being the non-metricity representation.  In the curvature representation, the torsion and non-metricity vanishes while in the teleparallel representation the curvature and non-metricity vanishes. Interestingly in the non-metricity representation, the curvature and torsion vanishes which eventually leads to the symmetric teleparallel gravity \cite{Nester99}. In this approach, the basic geometry of the gravitational action is represented by the non-metricity $Q$ of the metric. This has been further developed into the coincident gauge known as the $f(Q)$ gravity or non-metricity gravity \cite{Jimenez18}. Subsequently $f(Q)$ gravity has been extended to $f(Q,T)$ gravity \cite{Xu19} by including the trace of energy momentum tensor $T$. This modified gravity theory has been used to address some of the cosmology and astrophysics issues pertaining to early and late time evolution of the Universe. The late time cosmic acceleration issue has been studied in $f(Q,T)$ gravity  \cite{Pati21} and different bouncing scenarios have been discussed in \cite{Agrawal21}. The possible occurrence of future singularity has been studied in Ref. \cite{Pati22} whereas the dynamical system analysis has been performed in Ref. \cite{Pati23}. \\
 
The purpose of the present work is to construct possible Casimir wormholes in $f(Q,T)$ gravity and to study the effect of the GUP correction arising out of the minimal length concept in quantum mechanics.
The paper is organised as follows: in Sec. \ref{Sec.II}, a brief description of $f(Q,T)$ gravity and its mathematical formalism has been provided. In Sec. \ref{Sec.III}, we have discussed in brief the Casimir effect and the GUP correction and their application to traversable wormholes. Also we have studied the  effect of GUP correction on the wormhole geometry and other properties of traversable wormholes. The energy condition and exotic matter content are analysed in Sec. \ref{Sec.IV}. Finally the discussion and conclusion are presented in Sec. \ref{Sec.V}.

\section{Wormhole Taxonomy and $f(Q,T)$ gravity Field Equations} \label{Sec.II}
The action of $f(Q,T)$ gravity \cite{Xu19},
\begin{equation} 
S=\int\left[\dfrac{1}{16\pi}f(Q,T)+\mathcal{L}_{m}\right]d^{4}x\sqrt{-g},\label{eq.1}
\end{equation}
where, $f(Q,T)$ be the arbitrary function of the nonmetricity $Q$ and trace of energy momentum tensor $T$ and both can be respectively expressed as, $Q \equiv -g^{\mu \nu}( L^k_{~l\mu}L^l_{~\nu k}-L^k_{~lk}L^l_{~\mu \nu})$ and $T=g^{\mu \nu}T_{\mu \nu}$. The disformation tensor in $Q$ can be defined as, $L^k_{~l\gamma}\equiv-\frac{1}{2}g^{k\lambda}(\bigtriangledown_{\gamma}g_{l\lambda}+\bigtriangledown_{l}g_{\lambda \gamma}-\bigtriangledown_{\lambda}g_{l\gamma})$  . The determinant of the metric tensor and the matter Lagrangian respectively denoted as $g$ and $\mathcal{L}_m$ in the action. Now, varying the gravitational action \eqref{eq.1} with respect to the metric tensor, the field equations of $f(Q,T)$ gravity \cite{Xu19} can be obtained as, 
\begin{equation}
-\frac{2}{\sqrt{-g}}\bigtriangledown_{k}(f_{Q}\sqrt{-g}p^{k}_ {\mu \nu})-\frac{1}{2}fg_{\mu \nu}-f_{Q}(p_{\mu kl} Q^{\;\;\; kl}_{\nu}-2Q^{kl}_{\;\;\;\mu} p_{kl\nu})+f_{T}(T_{\mu \nu}+\Theta_{\mu \nu})=8 \pi T_{\mu \nu},\label{eq.2}
\end{equation}
where we denote, $f\equiv f(Q,T)$ and $f_Q=\frac{\partial f}{\partial Q}$. The super potential term is defined as, $p^{k}_{\mu \nu}=-\frac{1}{2}L^{k}_{\mu \nu}+\frac{1}{4}(Q^{k}-\tilde{Q}^{k})g_{\mu \mu}-\frac{1}{4}\delta^{k}_{(\mu}Q_{\nu)}$. The energy momentum tensor, $T_{\mu \nu}=\frac{-2}{\sqrt{-g}} \frac{\delta(\sqrt{-g}L_{m})}{\delta g^{\mu \nu}}$ and $\Theta_{\mu \nu}=g^{kl}\frac{\delta T_{kl}}{\delta g^{\mu \nu}}$. Also, The non-metricity tensor can be expressed as, $Q_{k}=Q_{k}^{\;\;\mu}\;_{\mu}$, $\tilde{Q}_{k}=Q^{\mu}\;_{k\mu}$.\\

The line element for the spherically symmetric wormhole in the coordinate ($t$, $r$, $\theta$, $\phi$) can be expressed as \cite{Morris88},  
\begin{equation}
ds^{2}=-e^{2\Phi(r)}dt^{2}+\frac{dr^{2}}{1-\frac{b(r)}{r}}+r^{2}(d\theta^{2}+sin^{2}\theta d\phi^{2}), \label{eq.3}
\end{equation} 
where $\Phi(r)$ and $b(r)$ are respectively the redshift and shape function. It is to note that the redshift function remains finite everywhere to avoid the presence of event horizons whereas the shape function needs to obey the flare-out condition of the throat. Now, using the a wormhole matter content and its stress-energy tensor, an anisotropic fluid source with a tensor that fulfills the energy criteria can be expressed as \cite{Varieschi16},
\begin{equation}
T_{\mu\nu}=(\rho+p_{t})U_{\mu}U_{\nu}+p_{t}g_{\mu\nu}+(p_{r}-p_{t})X_{\mu}X_{\nu},\label{eq.4}
\end{equation}
with $\rho$, $p_{r}$ and $p_{t}$ are respectively be the energy density, radial pressure and tangential pressure as measured in the fluid element in rest frame. Here $X_{\mu}$ is a space-like vector orthogonal to $U_{\mu}$, the four-velocity vector of the fluid satisfying the conditions $U_{\mu}U_{\nu}g^{\mu\nu}=-1$, $X_{\mu}X_{\nu}g^{\mu\nu}=1$ and the orthogonality condition, $U^{\mu}X_{\mu}= 0$. Hence, the diagonal form of the stress-energy tensor is, $T^{\mu}_{\nu}= diag[-\rho, p_r, p_t, p_t]$ . The experimental verification of all potential parameter combinations claimed that there is no significant difference in the wormhole solution whether $\Phi(r)\neq0$ or $\Phi(r)=0$. Therefore, for simplicity, we have opted for zero tidal force, i.e. $\Phi(r)=0$ and have derived the field equations of $f(Q,T)$ gravity for the line element \eqref{eq.3} as, 
\begin{eqnarray}
\frac{F}{r^2}\left[\frac{r-b(r)}{r}-b'(r)\right]+\frac{2\dot{F}(r-b(r))}{ r^2}+\frac{f}{2}&=&8\pi\rho, \label{eq.5} \\
\frac{F}{r^2}-\frac{f}{2}-\frac{2F(r-b(r))}{r^3}&=&p_{r}(8\pi+f_{T} )+\rho f_{T}, \label{eq.6} \\
\frac{-F(1-b'(r))}{2r^2}-\frac{F(r-b(r))}{2r^3}-\frac{\dot{F}(r-b(r))}{r^2}-\frac{f}{2}&=&p_{t}(8\pi+f_{T} )+\rho f_{T}, \label{eq.7}
\end{eqnarray}
where the notation, $F=\frac{\partial f}{\partial Q}$ and a prime denotes the derivative with respect to $r$, $f_{T}=\frac{\partial f}{\partial T}$. With an algebraic manipulations, the set of field equations \eqref{eq.5}-\eqref{eq.7} can be expressed as,

\begin{eqnarray}
\rho&=&\frac{1}{8\pi}\left[\frac{F}{r^2}\left(\frac{r-b(r)}{r}-b'(r)\right)+\frac{2\dot{F}(r-b(r))}{r^2}+\frac{f}{2}\right], \label{eq.8}\\ 
 p_{r}&=&\frac{1}{(8\pi+f_{T})}\left[\frac{F}{r^2}\left(\left(-1-\frac{f_{T}}{8\pi}\right)+\left(2+\frac{f_{T}}{8\pi}\right)\frac{b(r)}{r}+\frac{f_{T}}{8\pi}b'(r)\right)+\dot{F}\frac{f_{T}}{8\pi}\left(\frac{2b(r)}{r^{2}}-\frac{2}{r}\right)-\left(1+\frac{f_{T}}{8\pi}\right)\frac{f}{2}\right], \label{eq.9}\\
p_{t}&=&\frac{1}{(8\pi+f_{T})}\left[\frac{F}{r^2}\left(\left(-1-\frac{f_{T}}{8\pi}\right)+\left(\frac{1}{2}+\frac{f_{T}}{8\pi}\right)\frac{b(r)}{r}+\left(\frac{1}{2}+\frac{f_{T}}{8\pi}\right)b'(r)\right)\right] \nonumber\\
&+&\frac{\dot{F}}{8\pi+f_T}\left[\left(\left(-1-\frac{f_{T}}{4\pi}\right)\frac{1}{r}+\left(1+\frac{f_{T}}{4\pi}\right)\frac{b(r)}{r^2}\right)-\left(1+\frac{f_{T}}{8\pi}\right)\frac{f}{2}\right]. \label{eq.10}
\end{eqnarray}

In their seminal work, Xu et al. \cite{Xu19} have suggested three forms of the function $f(Q,T)$ such as, (i) $f(Q,T)=\lambda_1 Q+\lambda_2 T$, (ii) $f(Q,T)=\lambda_1 Q^{n+1}+\lambda_2 T$ and (iii) $f(Q,T)=-\lambda_1 Q-\lambda_2 T^2$, where $\lambda_1$ and $\lambda_2$ are two arbitrarily constants. However, here we consider the first case $f(Q,T)=\lambda_1 Q+\lambda_2 T$ to obtain the wormhole geometry solution such that, $F=\lambda_1$ and $\dot{F}=0$. With this consideration, the nonmetricity becomes, 
\begin{equation}
Q=-\frac{2}{r}\left(1-\frac{b(r)}{r}\right)\frac{1}{r} .  \label{eq.11} 
\end{equation} 
Now, the set of field equations \eqref{eq.8}-\eqref{eq.10} reduce to,
\begin{eqnarray}
\rho&=&\frac{1}{8\pi}\left[\frac{\lambda_1}{r^2}\left(\frac{r-b(r)}{r}-b'(r)\right)+\frac{f}{2}\right], \label{eq.12}\\
p_{r}&=&\frac{1}{(8\pi+\lambda_2)}\left[\frac{\lambda_1}{r^2}\left(\left(-1-\frac{\lambda_2}{8\pi}\right)+\left(2+\frac{\lambda_2}{8\pi}\right)\frac{b(r)}{r}+\frac{\lambda_2}{8\pi}b'(r)\right)-\left(1+\frac{\lambda_2}{8\pi}\right)\frac{f}{2}\right],\label{eq.13}\\
p_{t}&=&\frac{1}{(8\pi+\lambda_2)}\left[\frac{\lambda_1}{r^2}\left(\left(-1-\frac{\lambda_2}{8\pi}\right)+\left(\frac{1}{2}+\frac{\lambda_2}{8\pi}\right)\frac{b(r)}{r}+\left(\frac{1}{2}+\frac{\lambda_2}{8\pi}\right)b'(r)\right)-\left(1+\frac{\lambda_2}{8\pi}\right)\frac{f}{2}\right]. \label{eq.14}
\end{eqnarray}
From Eqns. \eqref{eq.12}-\eqref{eq.14}, the trace of the energy momentum tensor can be obtained as,
\begin{equation}
T=-\rho+p_{r}+2p_{t} =\frac{2 \lambda_1 b'(r)}{(\lambda_2+8 \pi ) r^2}. \label{eq.15}
\end{equation}
We have the trace of nonmetricty tensor and energy momentum tensor, so we can substitute this in $f\equiv f(Q,T)=\lambda_1 Q+\lambda_2 T$ in Eqns. \eqref{eq.12}-\eqref{eq.14} to obtain,  
\begin{eqnarray}
\rho&=&-\frac{\lambda_1 b'(r)}{(\lambda_2+8 \pi ) r^2},\label{eq.16}\\
p_{r}&=&\frac{\lambda_1 b(r)}{r^3(\lambda_2+8 \pi )},\label{eq.17}\\
p_{t}&=&\frac{\lambda_1 (-b(r)+rb'(r))}{2r^3 (\lambda_2+8 \pi ) }.\label{eq.18}
\end{eqnarray}
 
To note, when we substitute $\lambda_2=0$, the above set of equations \eqref{eq.16}-\eqref{eq.18} reduce to that of the $f(Q)$ gravity. \\
 
In classical relativity, the energy conditions are violated in wormhole geometry, however it may have different behaviour in the modified theories of gravity e.g. $f(Q,T)$ gravity. The Raychaudhuri equation  ~\cite{Raychaudhuri1955,Raychaudhuri1957a,Raychaudhuri1957b} describes the energy conditions in terms of time-like and space-like curves\cite{Hawking99}.  The energy conditions in terms of radial and tangential pressure for the traversable wormhole within the $f(Q,T)$ gravity are expressed as,

\begin{eqnarray}
\rho+p_{r}&=&\frac{\lambda_1 (b(r)-r b'(r))}{r^3(8 \pi+\lambda_2 )}, \nonumber \\
\rho+p_{t}&=&-\frac{\lambda_1 (b(r)+r b'(r))}{2 r^3(8 \pi+\lambda_2 ) },\nonumber \\
\rho-p_{r}&=&-\frac{\lambda_1 (b(r)+r b'(r))}{ r^3(8 \pi+\lambda_2 )}, \nonumber \\
\rho-p_{t}&=&\frac{\lambda_1 (b(r)-3r b'(r))}{2r^3 (8 \pi+\lambda_2 )}, \nonumber\\
p_{t}-p_{r}&=&\frac{\lambda_1 (-3 b(r)+r b'(r))}{2r^3(8 \pi+\lambda_2 ) },\nonumber \\
\rho+p_{r}+2p_{t}&=&0. \end{eqnarray}

\section{Casimir Wormholes and GUP correction}\label{Sec.III}
In classical GR, though traversable wormhole solutions are very much possible, but its very existence as well as stability depend on the invoked amount of exotic matter content. Particularly, the NEC of the wormhole matter field should be violated which requires a negative energy density source to keep open the mouth of the wormhole for any physical object to pass through its tunnel. Within the classical regime, it is not possible to have a negative energy source and therefore, classical traversable wormhole may not exist in nature. However, quantum mechanics provides us some opportunity to explore certain sources with negative energy density. One such aspect is the Casimir energy as predicted long back by H. Casimir \cite{Casimir48} and confirmed through experiments later on by others \cite{Sparnaay57,Mohideen1998,Bressi2002,Vezzoli2019,Wilczek2020}. The Casimir effect involves the existence of an attractive force between two parallel, conducting and uncharged conductors resulting from the distortion of vacuum of the electromagnetic field. The zero-point energy of the quantum electrodynamics distorted by the plates has some bearing on the negative Casimir energy. In fact, the quantum field fluctuation leading to a negative energy density which represents the only source of exotic matter that can be produced within a laboratory condition \cite{Garattini21}.

The attractive Casimir force develops between the plates because of the renormalized negative energy
\begin{equation}
E(d)=-\frac{\pi^2}{720}\frac{A}{d^3},
\end{equation}
where $A$ and $d$ respectively denote the plate surface area and the plate separation distance. It is obvious that the Casimir energy decreases if the plates are moved closer to each other. 

Now the Casimir energy density (Casimir energy per unit volume $\rho=\frac{E}{V}$) may be obtained as
\begin{equation}
\rho(d)=-\frac{\pi^2}{720}\frac{1}{d^4},\label{eq:rho}
\end{equation}
and consequently the pressure becomes
\begin{equation}
p(d)=-\frac{\pi^2}{240}\frac{1}{d^4}.\label{eq:p}
\end{equation}

Another important aspect in quantum mechanics is the existence of minimal length scale of the order of Planck length $l_p=\sqrt{\frac{G\hbar}{c^3}}\simeq 10^{-35}m$  that limits the resolution of small distances in the spacetime \cite{Pedram2012}. Such a minimal length scale is implied in many quantum gravity theories \cite{Veneziano89, Konishi90, Maggiore93, Witten96, Garay95, Maggiore93a, Magueijo02, Cortes05, Kempf95, Nozari12}. It is to be noted here that, the spatial resolution can not be improved below this characterisitc length scale which obviously demands a corresponding generalization of the position-momentum uncertainty principle (GUP) in the form $
\triangle x\triangle p \geq \frac{1}{2}\left[1+\beta\left(\triangle p\right)^2\right]$, where $\beta$ represents the parameter of the GUP correction. In fact, the minimal length concept in quantum mechanics with the GUP depends upon the maximally localized quantum states. In momentum representation, the maximally localized states are given by
\begin{equation}
\psi ^{ML}= \frac{1}{(2\pi)^{3/2}} \Omega(p) \exp^{-i \left[k.r-w(p)t\right]},
\end{equation}
which satisfies the equation
\begin{equation}
\left[\hat{x}-\langle x \rangle+\frac{\langle [\hat{x}, \hat{x}] \rangle}{2(\bigtriangleup p)^{2} (\hat{p}-\langle p\rangle)} \right] |\psi\rangle =0.
\end{equation}
The commutation relation after incorporating the correction due to the minimal length concept can be generalised to $n$ dimension as \cite{Frassino12}
\begin{equation}
\left[\hat{x}_i, \hat{p}_j \right]= i \left[f(\hat{p}^2)\delta_{ij}+g(\hat{p}^2) \hat{p}_i \hat{p}_j \right],
\end{equation}
where, $i,j=1,.......,n,$
and $f(\hat{p}^2)$ and $g(\hat{p}^2)$ are the generic functions.

From the the translational and rotational invariance of the generalized uncertainty principle, one may get the generic functions. Use of different generating functions leads to different construction of the mazximally localized quantum states. In this work, we would like to focus upon two such constructions of the maximally localized quantum states by two different groups of scientists, viz. Kempf, Mangano and Mann (KMM) \cite{Kempf95} and Detournay, Gabriel and Spindel (DGS) \cite{Detournay02}. Through a detailed calculation of the Hamiltonian and the corrections to the Casimir energy due to the minimal length driven GUP keeping upto first order in $\beta$, Frassino and Panella \cite{Frassino12} obtained the Casimir energy density as
\begin{eqnarray}
\rho_i(a) &=& -\frac{\pi^2}{720 }\frac{1}{d^4}\left[1+\xi_{i}\frac{\beta}{d^2}\right],
\end{eqnarray}
where 
\begin{eqnarray}
\xi_{KMM} &=& \pi^2\left(\frac{28+3\sqrt{10}}{14}\right),\\
\xi_{DGS} &=& 4\pi^2\left(\frac{3+\pi^2}{21}\right).
\end{eqnarray}

One should note that,  $\frac{\xi_{KMM}}{\xi_{DGS}}=1.0923$ and therefore the contribution coming from the GUP correction term may decrease by a factor of $1.0923$ for the DGS construction \cite{Detournay02} as compared to that of KMM construction \cite{Kempf95}.

We may now replace the plate separation distance $d$ by the radial coordinate $r$ and integrate the field equations in the extended symmetric teleparallel gravity to obtain the shape function for the GUP corrected Casimir wormhole as
\begin{equation}
    b(r,\beta)=r_0-k\left(\frac{1}{r}-\frac{1}{r_0}\right)-\frac{k\xi_i\beta}{3}\left(\frac{1}{r^3}-\frac{1}{r_0^3}\right),
\end{equation}
where $k=\frac{\pi^2}{720}\left(\frac{\lambda_2+8\pi}{\lambda_1}\right)$.
It is obvious that the above shape function reduces to the wormhole throat for $r=r_0$.
The first derivative of the shape function is
\begin{equation}
    b^{\prime}(r,\beta)=\frac{k}{r^2}\left(1+\frac{\xi_i\beta}{r^2}\right),
\end{equation}
which becomes 
\begin{equation}
    b^{\prime}(r_0,\beta)=\frac{k}{r_0^2}\left(1+\frac{\xi_i\beta}{r_0^2}\right),
\end{equation}
at the wormhole throat. The geometry modification of the gravity theory affects the wormhole geometry through the quantity $k$ and appear in the second and third terms of the expression of the shape function. However, the GUP correction appears only in the third term of the shape function. In the absence of the GUP correction, the shape will reduce to the usual case of Casimir wormhole 
\begin{equation}
    b(r)=r_0-k\left(\frac{1}{r}-\frac{1}{r_0}\right).
\end{equation}

\begin{figure}[H]
    \centering
    \includegraphics[scale=.325]{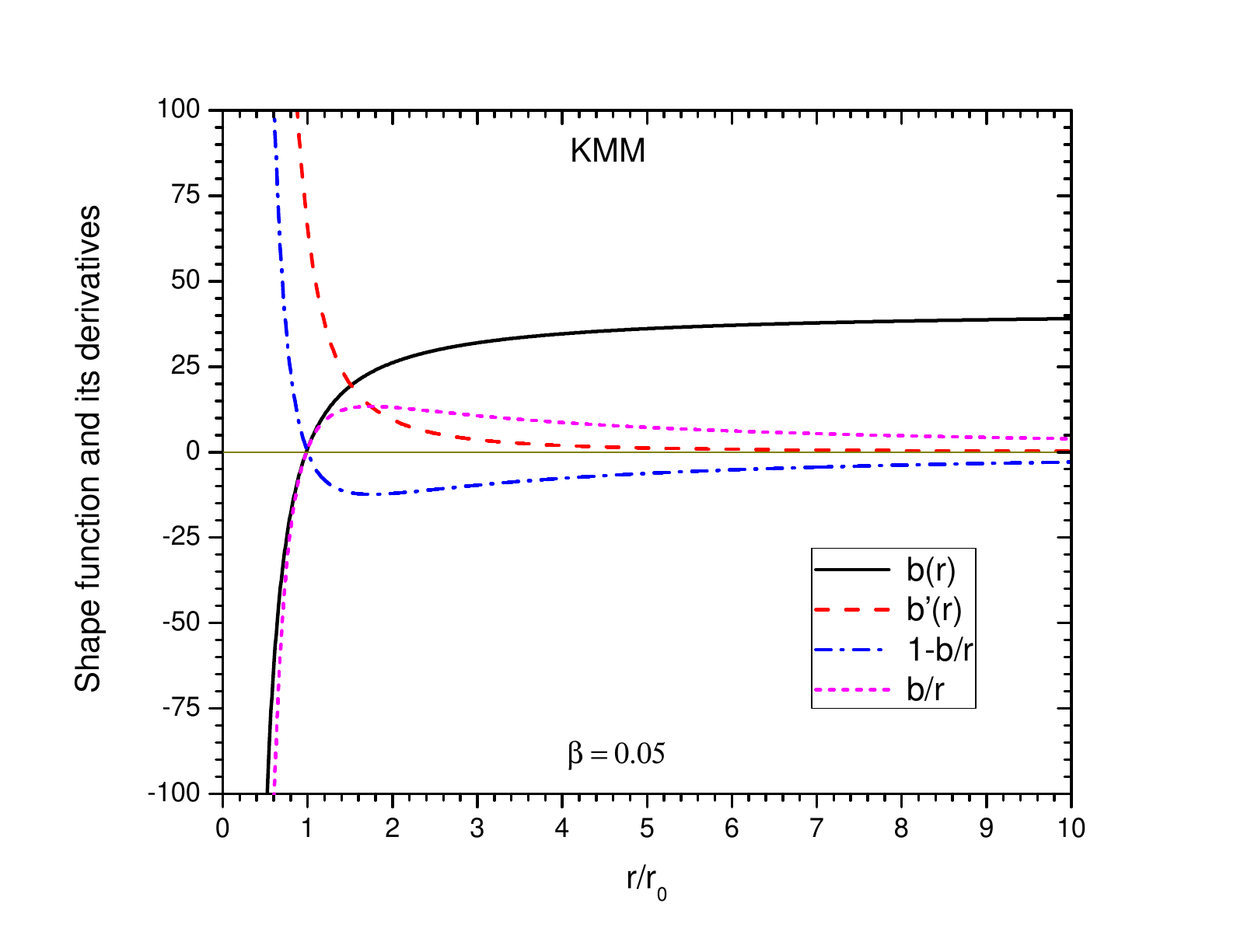}
    \includegraphics[scale=.325]{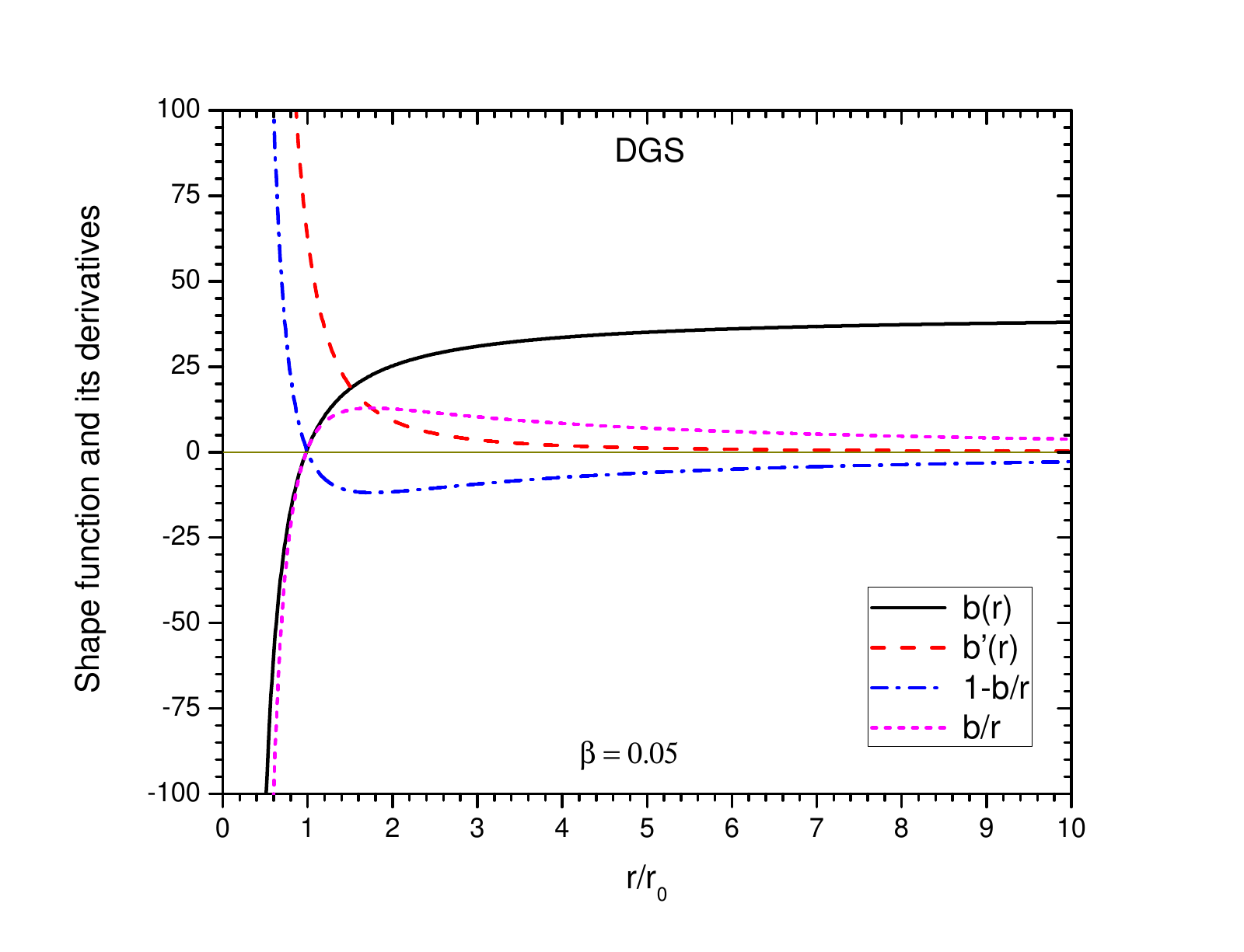}
    \caption{ The graphical bahaviour of the shape function and its derivatives for the GUP corrected Casimir wormhole (a) with KMM construction  and (b) with DGS construction for $\beta = 0.05$. We have considered  the model parameters as $\lambda_{1}= -4.4$ and  $\lambda_{2}= 0.01.$}
    \label{fig:1}
\end{figure}

\begin{figure}[H]
    \centering
    \includegraphics[scale=.325]{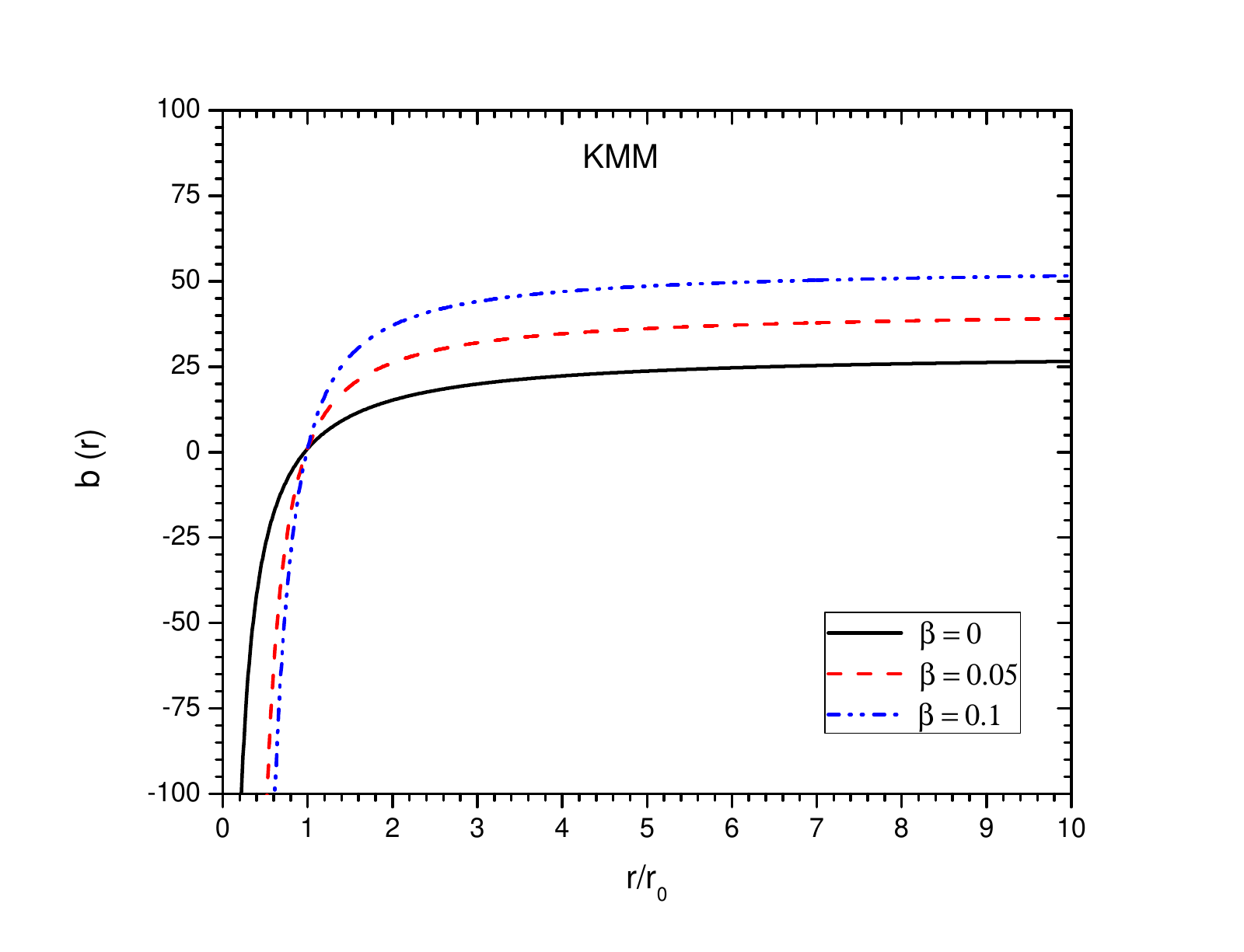}
     \includegraphics[scale=.325]{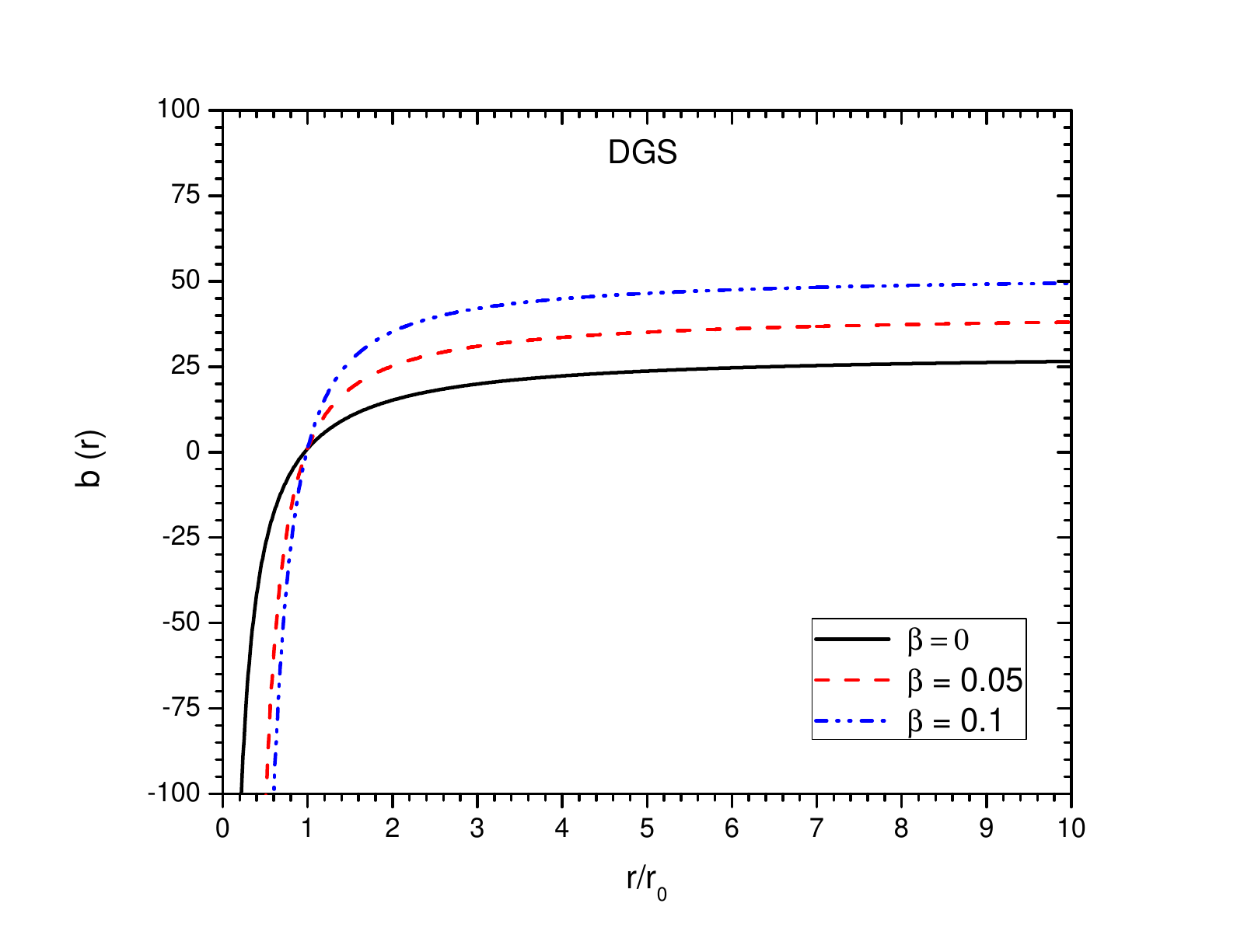}
    \caption{ The graphical bahaviour of shape function for the GUP corrected Casimir wormhole (a) with  KMM construction  and (b) with DGS construction for different values of $\beta$. We have considered  the model parameters as $\lambda_{1}= -4.4$ and  $\lambda_{2}= 0.01.$}
    \label{fig:2}
\end{figure}

In FIG. \ref{fig:1}, we show the shape function of the GUP corrected Casimir wormhole and other related functions such as $b^{\prime}(r,\beta), b/r$ and  $1-b/r$ as functions of the reduced radial distance $r/r_0$. For the sake of conveninec, we consider the wormhole throat radius to be  $r_0 =1$. The parameter space used in the present work is $\beta=0.05, \lambda_1=-4.4$ and $\lambda_2=0.01$.  In order to draw a comparison between different choices of the construction of the maximally localized quantum states, we have considered two specific choices denoted as KMM and DGS constructions. The shape function for the Casimir wormhole as obtained for these two constructions are shown in FIG.\ref{fig:1}. In both the cases, the only difference is the value of the parameter $\xi_i$ which is greater for the KMM construction and therefore, the value of the shape function for KMM construction is slightly higher as compared to that of the DGS construction. In both the cases, the shape function satisfies the flare out condition and reduces to $r_0$ at the wormhole throat. In order to assess the role of the GUP parameter on the shape function and the wormhole geometry, we have shown the Casimir wormhole shape function for different values of the GUP parameter $\beta$ namely $\beta=0, 0.05$ and 0.1 in FIG.\ref{fig:2}. The curve corresponding to $\beta=0$ represents the Casimir wormhole without GUP correction taken into account. It is obvious that, the GUP parameter has an exemplified effect outside the wormhole throat. With an increase in $\beta$, the shape function is found to increase substantially. However, within the wormhole throat, the GUP parameter has a role to decrease the value of the shape function at a given radial distance. 

%%%%%%%%%%%%%%%%%%%%%%%%%%%%%%%%%%%%%%%%%%%%%%%%%%%%%%%%%%%%%%%%%%%%%%%%%%%%%%%%%%%%%%%%%%%%%%%%%%%%%%%%%%%%%%%%%%%%%%%%%%%%%%%%%%%%%%%%%%%%%%%%%%%%%%%%%%%%%%%

%%%%%%%%%%%%%%%%%%%%%%%%%%%%%%%%%%%%%%%%%%%%%%%%%%%%%%%%%%%%%%%%%%%%%%%%%%%%%%%%%%%%%%%%%%%%%%%%%%%%%%%%%%%%%%%%%%%%%%%%%%%%%%%%%%%%%%%%%%%%%%%%%%%%%%%%%%%%%%%

The radial and tangential pressures for the GUP corrected Casimir wormhole are obtained respectively as 
\begin{equation}
    p_r (r, \beta) =\frac{\lambda_1}{\lambda_2+8\pi}\frac{1}{r^3}\left[r_0-k\left(\frac{1}{r}-\frac{1}{r_0}\right)-\frac{k\xi_i\beta}{3}\left(\frac{1}{r^3}-\frac{1}{r_0^3}\right)\right],
\end{equation}

\begin{equation}
    p_t (r, \beta)=-\frac{\lambda_1}{\lambda_2+8\pi}\frac{1}{r^3}\left[\frac{r_0}{2}-k\left(\frac{1}{r}-\frac{1}{2r_0}\right)+\frac{k\xi_i\beta}{3}\left(\frac{1}{2r_0^3}-\frac{2}{r^3}\right)\right].
\end{equation}

One should note that, the Casimir effect is visible in the second and third terms of the respective expressions of the radial pressure and tangential pressure. However, the GUP correction appears only in the respective third terms. From the expressions of the radial and tangential pressures, we may define the radial and tangential equation of state (EoS) parameters, respectively, as  
\begin{eqnarray}
\omega_r (r, \beta) &=& \frac{p_r(r, \beta)}{\rho}=-\frac{3r_0^4r^3-3kr_0^3r^2+3kr^3r_0^2-k\xi_i \beta r_0^3+k\xi_i \beta r^3}{3kr_0^3(r^2+\xi_i \beta)},\\
\omega_t (r, \beta) &=& \frac{p_t(r, \beta)}{\rho}=\frac{3r_0^4r^3+3kr^3r_0^2-6kr_0^3r^2-4kr_0^3\xi_i \beta+kr^3\xi_i \beta}{6kr_0^3(r^2+\xi_i \beta)},
\end{eqnarray}
where $\rho=-\frac{\pi^2}{720}\frac{1}{r^4}\left[1+\frac{\xi_i\beta}{r^2}\right]$.

The anisotropy in the pressure of the exotic matter of the wormhole becomes
\begin{equation}
\bigtriangleup \omega(r, \beta)=\frac{\omega_t(r, \beta)}{\omega_r(r, \beta)}=\frac{6kr_0^3r^2-3r_0^4r^3-3kr^3r_0^2-kr^3\xi_i\beta+4kr_0^3\xi_i \beta}{2(3r^3r_0^4-3kr_0^3r^2+3kr^3r_0^2-kr_0^3\xi_i \beta+kr^3\xi_i \beta)}.
\end{equation}

We may also define the pressure anisotropy through 
\begin{equation}
    \triangle p=p_t-p_r=\frac{\lambda_1}{\lambda_2+8\pi}\frac{1}{r^3}\left[-\frac{3r_0}{2}+k\left(\frac{2}{r}-\frac{3}{2r_0}\right)+\frac{k\xi_i\beta}{3}\left(\frac{1}{r^3}-\frac{1}{2r_0^3}\right)\right].
\end{equation}

It is interesting to note that, the GUP modification to the Casimir energy affects the wormhole pressure both in the radial and tangential directions but at the throat, the magnitude of anisotropy in the pressure remains the same as that without modification. Beyond the wormhole radius, the anisotropy factor decreases with the increase in $\beta$. However, for a radial distance less than the throat $\bigtriangleup\omega(r, \beta)$ increases with $\beta$. In comparison to the usual Casimir wormholes, the behaviour of $\bigtriangleup \omega (r,\beta)$ is quite different at a radial distance $r\textless r_0 $. At the wormhole throat, the pressure anisotropy parameter becomes independent of the GUP correction parameter $\beta$.

\begin{figure}[H]
    \centering
    \includegraphics[scale=.325]{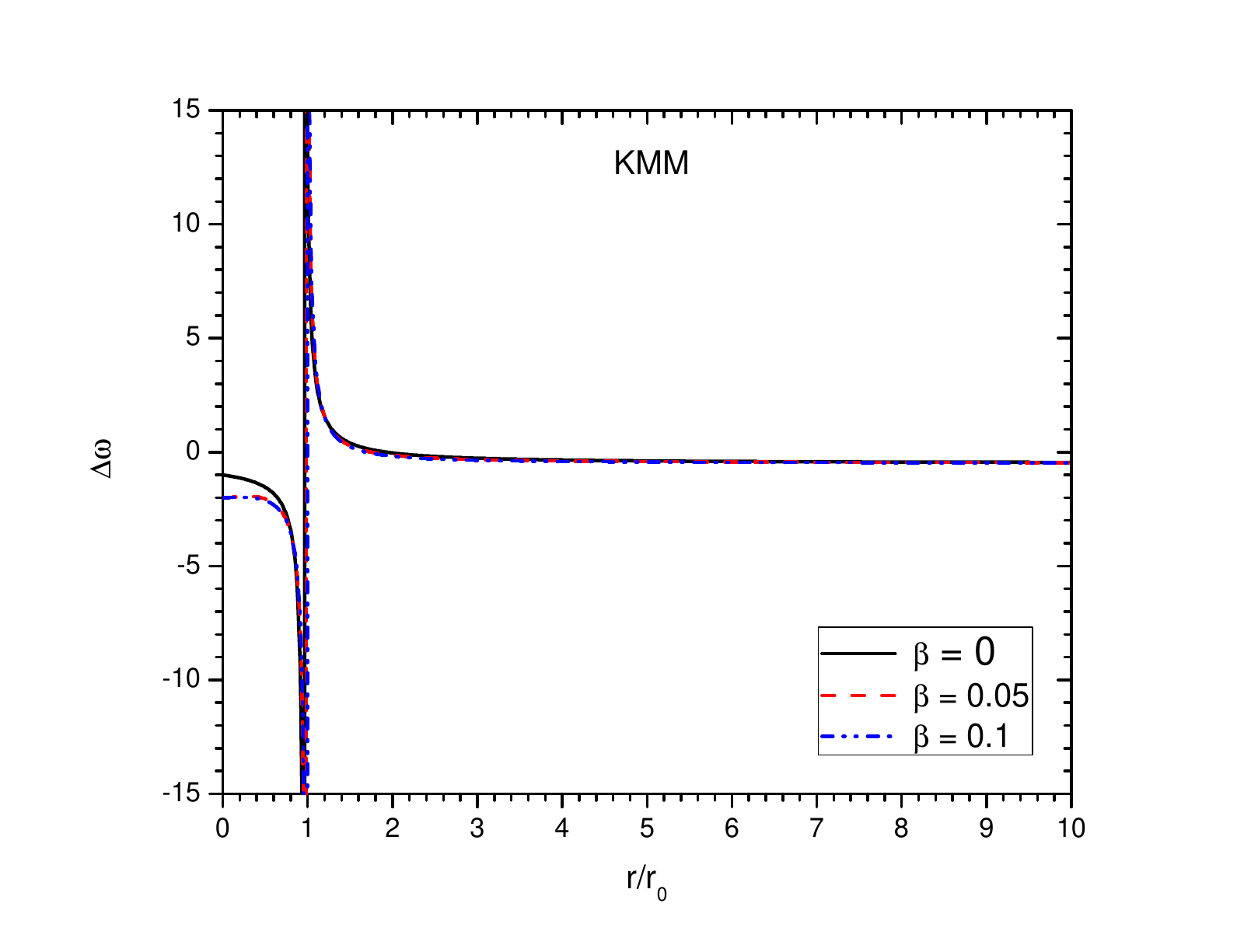}
     \includegraphics[scale=.325]{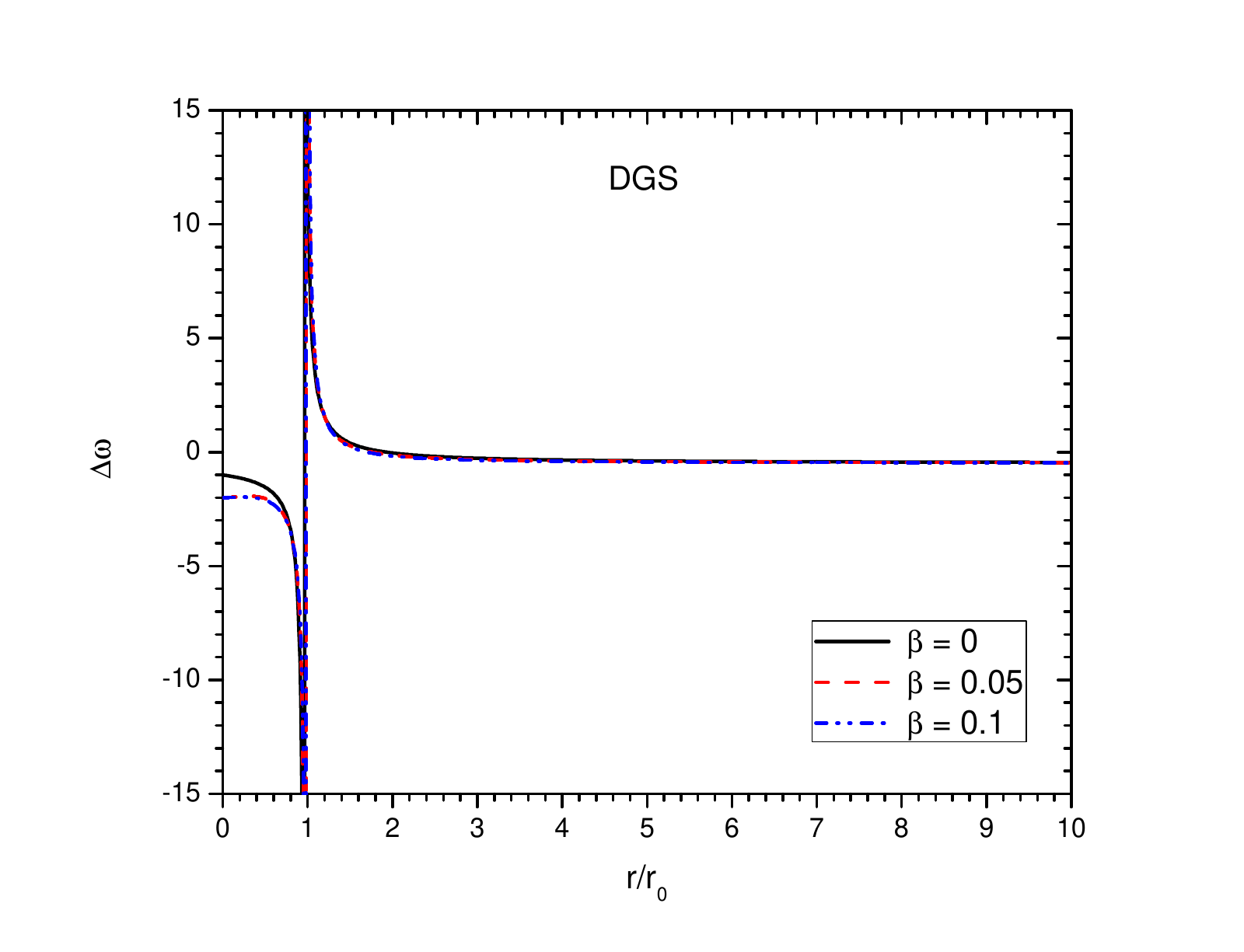}
    \caption{ The graphical bahaviour of pressure anisotropy $\bigtriangleup \omega$ for the GUP corrected Casimir wormhole (a) with  KMM construction  and (b) with DGS construction for different values of $\beta$. We have considered  the model parameters as $\lambda_{1}= -4.4$ and  $\lambda_{2}= 0.01.$}
    \label{fig:3}
\end{figure}

\begin{figure}[H]
    \centering
    \includegraphics[scale=.325]{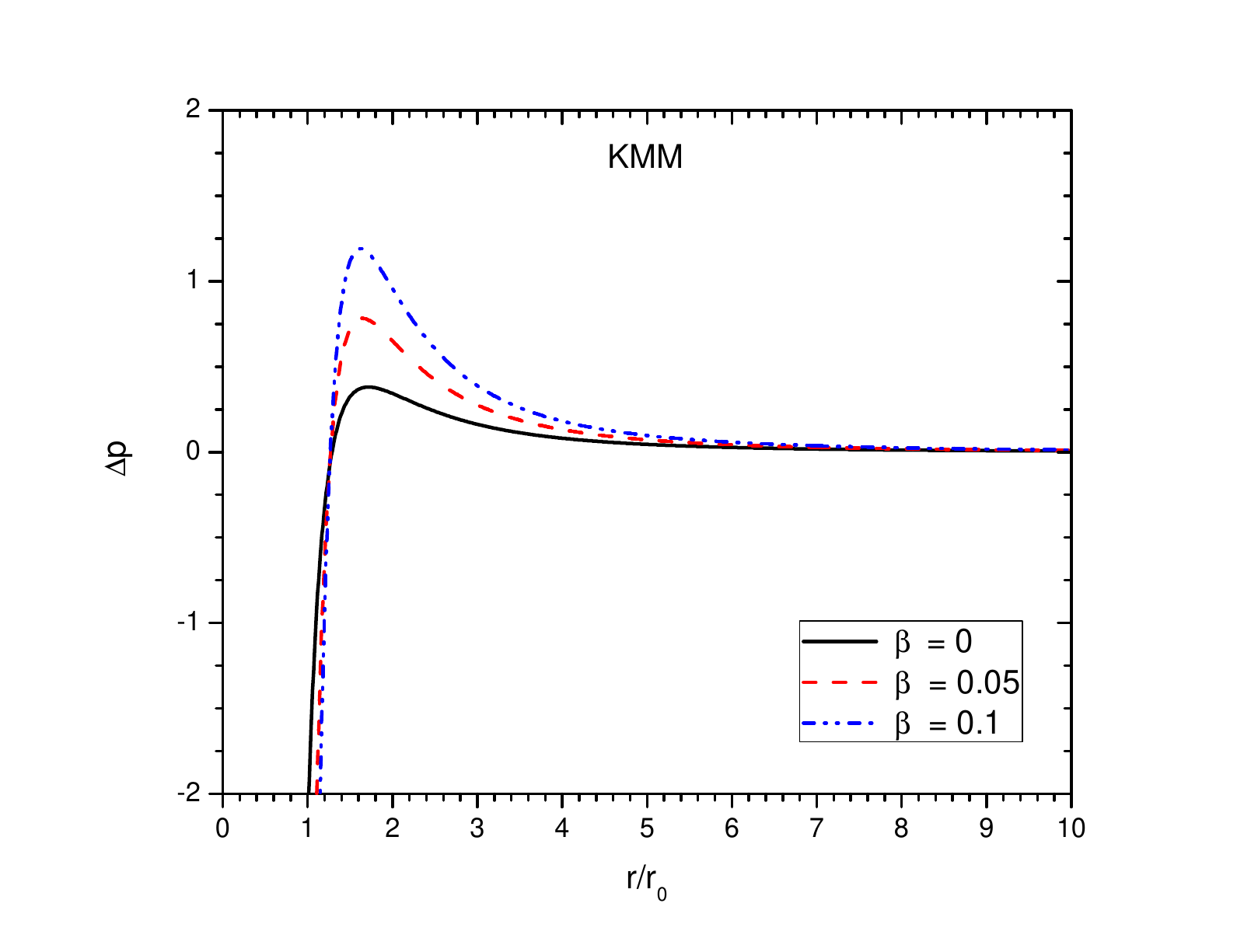}
     \includegraphics[scale=.325]{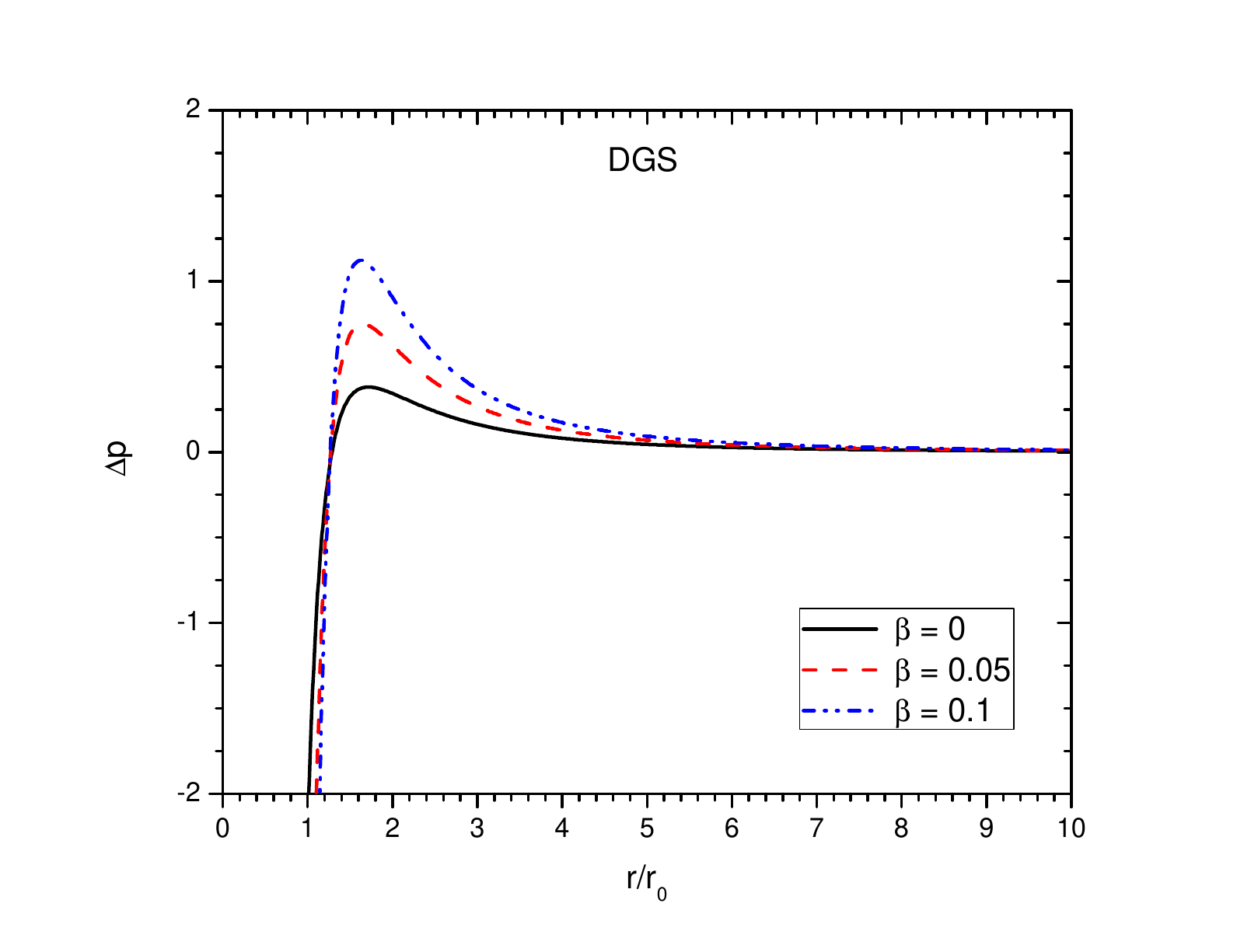}
    \caption{ The graphical bahaviour of $\triangle p$ for the GUP corrected Casimir wormhole (a) with  KMM construction  and (b) with DGS construction for different values of $\beta$. We have considered  the model parameters as $\lambda_{1}= -4.4$ and  $\lambda_{2}= 0.01.$}
    \label{fig:4}
\end{figure}

In FIG.\ref{fig:3} the pressure anisotropy as defined through the radial and tangential EoS parameter are shown for different values of the GUP parameter. In the left panel of the figure, we plot for the KMM construction and in the right panel that for the DGS construction. As we have already stated, the KMM and DGS constructions provide similar results but differ only in numerical values of $\bigtriangleup \omega(r, \beta)$. One should note that, the GUP parameter affects the $\bigtriangleup \omega(r, \beta)$ only near the wormhole throat. Away from the throat, the pressure anisotropy almost vanishes and therefore, the GUP parameter has least affect upon it. In FIG. \ref{fig:4}, the plot of $\triangle P(r, \beta)$ for both the constructions are shown. In general, the pressure anisotropy in the exotic fluid content of the Casimir wormhole increases upto a radial distance twice the throat radius and after it subsides and vanishes at far distance from the throat. The GUP parameter greatly exemplifies the pressure anisotrpy within $r\simeq 2r_0$. However, the influence of $\beta$ decreases as we move away from the throat. 

%%%%%%%%%%%%%%%%%%%%%%%%%%%%%%%%%%%%%%%%%%%%%%%%%%%%%%%%%%%%%%%%%%%%%%%%%%%%%%%%%%%%%%%%%%%%%%%%%%%%%%%%%%%%%%%%%%%%%%%%%%%%%%%%%%%%%%%%%%%%%%%%%%%%%%%%%%%%%%%

%%%%%%%%%%%%%%%%%%%%%%%%%%%%%%%%%%%%%%%%%%%%%%%%%%%%%%%%%%%%%%%%%%%%%%%%%%%%%%%%%%%%%%%%%%%%%%%%%%%%%%%%%%%%%%%%%%%%%%%%%%%%%%%%%%%%%%%%%%%%%%%%%%%%%%%%%%%%%%%

\begin{figure}[H]
    \centering
    \includegraphics[scale=.450]{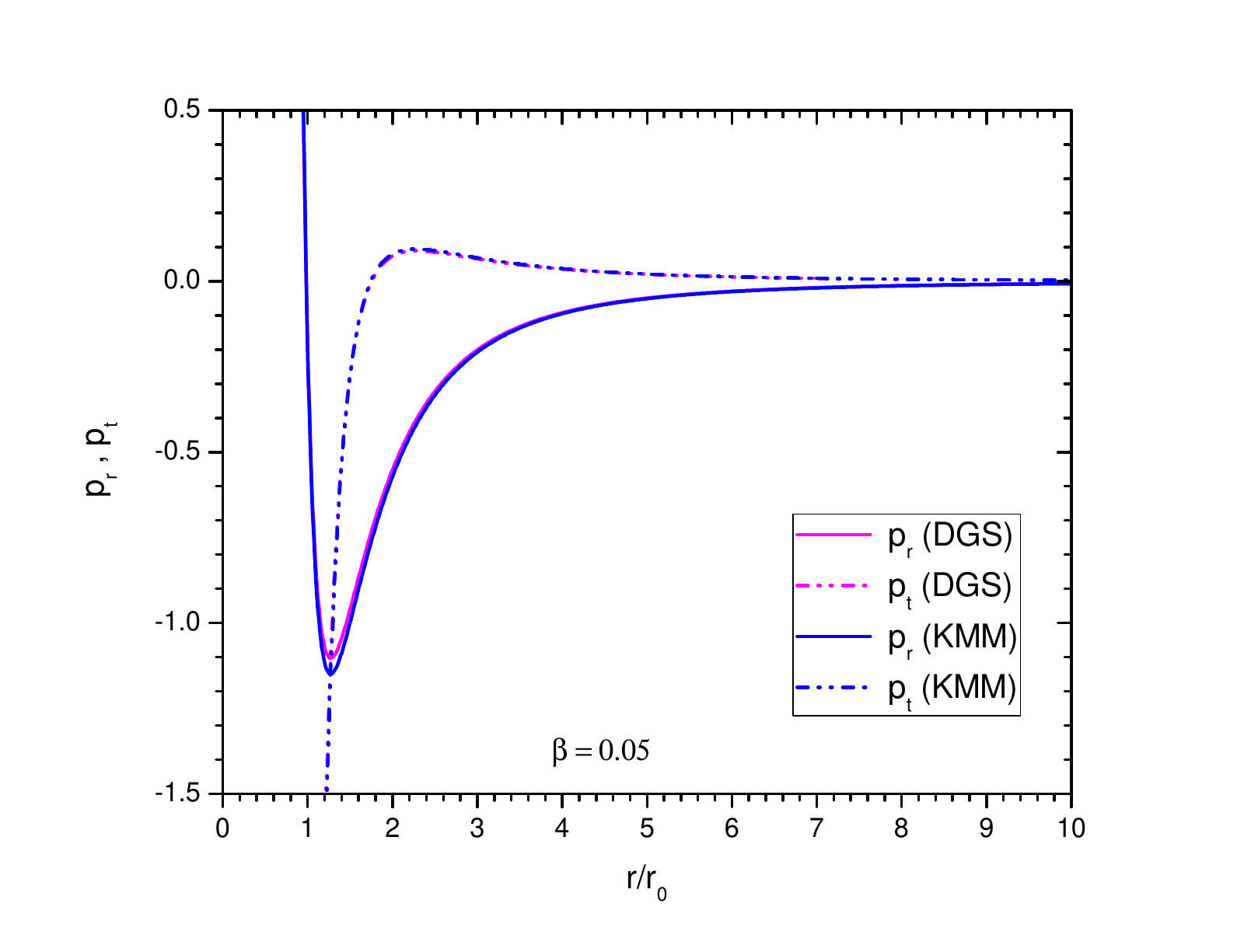}
    \caption{ The graphical bahaviour of the $p_r, p_t$and its derivatives for the GUP corrected Casimir wormhole with DGS construction for $\beta = 0.05$. We have considered  the model parameters as $\lambda_{1}= -4.4$ and  $\lambda_{2}= 0.01.$}
    \label{fig:5}
\end{figure}

The behaviour of $\bigtriangleup \omega(r, \beta)$ altogether changes 
 near the wormhole throat, i.e., around $r\simeq r_0$. At this radial distance, $\bigtriangleup \omega(r, \beta)$ suddenly changes its sign. In order to understand this behaviour, we have plotted the radial and tangential pressures for both the constructions for a given $\beta$ in FIG. \ref{fig:5}. One may observe that, the radial pressure decreases from a positive value to attain a negative minimum and then rises with the radial distance to vanish at large distnaces. On the other hand, the tangential pressure increases from a negative value to attain a maximum in the positive domain and then decrease to null values. During these evolution, near the wormhole throat, the radial and  tangential pressures both become negative so that there occurs a sudden change in sign in the behaviour of $\bigtriangleup \omega(r, \beta)$.

%%%%%%%%%%%%%%%%%%%%%%%%%%%%%%%%%%%%%%%%%%%%%%%%%%%%%%%%%%%%%%%%%%%%%%%%%%%%%%%%%%%%%%%%%%%%%%%%%%%%%%%%%%%%%%%%%%%%%%%%%%%%%%%%%%%%%%%%%%%%%%%%%%%%%%%%%%%%%%%
%%%%%%%%%%%%%%%%%%%%%%%%%%%%%%%%%%%%%%%%%%%%%%%%%%%%%%%%%%%%%%%%%%%%%%%%%%%%%%%%%%%%%%%%%%%%%%%%%%%%%%%%%%%%%%%%%%%%%%%%%%%%%%%%%%%%%%%%%%%%%%%%%%%%%%%%%%%%%%%

\section{Energy Conditions and Exotic Matter content} \label{Sec.IV}
In general, due to the presence of exotic matter in wormholes, some energy conditions are violated. Particularly the NEC defined as $t_{\mu \nu}k^{\mu}k^{\nu}\geq 0$ or $\rho(r,\beta)+p_r(r,\beta)\geq 0$ is violated. In this Section, we intend to check whether the GUP corrected Casimir wormholes in the $f(Q,T)$ gravity theory satisfy the energy conditions. The NEC for the GUP corrected Casimir wormholes can be assessed from the expression of NEC\\

\textbf{NEC:}
\begin{equation}
\rho(r, \beta)+p_r(r, \beta)= \frac{\lambda_1}{\lambda_2+8\pi} \left[\frac{-2k}{r^{4}}+\frac{r_0}{r^{3}}+\frac{k}{r_0 r^{3}}-\frac{4k\xi_i \beta}{3r^{6}}+\frac{k\xi_i \beta}{3r^{3}r_0^{3}} \right].
\end{equation}
 
In the above, the GUP correction term is proportional to the minimal uncertainty parameter $\beta$ . For a radial  distance $r<r_0$, obviously the right hand side of the above equation is a negative quantity and therefore the NEC is violated. With an increase in $\beta$, the contribution becomes more and more negative. The role of $\lambda$ is to minimise the NEC violation through the factor $b_1$. In the limit $\beta \rightarrow 0$, the above equation reduces to that of a Casimir wormhole:
\begin{equation}
\rho(r)+p_r(r)=\frac{\lambda_1}{\lambda_2+8\pi} \left[\frac{-2k}{r^{4}}+\frac{r_0}{r^{3}}+\frac{k}{r_0 r^{3}} \right].
\end{equation}
At the throat, the NEC reduces to
\begin{equation}
\rho(r_0,\beta)+p_r(r_0,\beta)=\frac{\lambda_1}{\lambda_2+8\pi} \left[\frac{1}{r_0^{2}} \left(1-\frac{k}{r_0^{2}}-\frac{k\xi_i\beta}{r_0^{4}}\right) \right].
\end{equation}
Since, the right side of the above equation is a negative quantity, it is obvious that, the NEC is violated by the GUP corrected Casimir wormhole at the throat.

The strong energy condition(SEC) is given by $\rho(r,\beta)+2p_t(r,\beta)\geq 0$. Another way to express the SEC is $\rho(r,\beta)+p_r(r,\beta)+2p_t(r,\beta)\geq 0$. For this statement, we have
\begin{equation}
\rho(r,\beta)+p_r(r,\beta)+2p_t(r,\beta)=0
\end{equation}
and other energy conditions are
\begin{eqnarray}
\rho(r)-p_r(r) &=& \frac{\lambda_1}{\lambda_2+8\pi} \left[\frac{-r_0}{r^{3}}-\frac{k}{r_0 r^{3}}-\frac{2k\xi_i \beta}{3r^{6}}-\frac{k\xi_i \beta}{3r^{3}r_0^{3}} \right]\\
\rho(r_0)-p_r(r_0) &=& \frac{\lambda_1}{\lambda_2+8\pi} \left[\frac{-1}{r_0^{2}}-\frac{k}{r_0^{4}}-\frac{k\xi_i \beta}{r_0^{6}} \right]\\
\rho(r)+p_t(r) &=& \frac{\lambda_1}{\lambda_2+8\pi} \left[\frac{-r_0}{2r^{3}}-\frac{k}{2r_0 r^{3}}-\frac{k\xi_i \beta}{3r^{6}}-\frac{k\xi_i \beta}{6r^{3}r_0^{3}} \right]\\
\rho(r_0)+p_t(r_0) &=& \frac{\lambda_1}{\lambda_2+8\pi} \left[\frac{-1}{2r_0^{2}}-\frac{k}{2r_0^{4}}-\frac{k\xi_i \beta}{2r_0^{6}} \right]\\
\rho(r)-p_t(r) &=& \frac{\lambda_1}{\lambda_2+8\pi} \left[\frac{r_0}{2r^{3}}-\frac{2k}{r^{4}}+\frac{k}{2r_0 r^{3}}-\frac{5k\xi_i \beta}{3r^{6}}-\frac{k\xi_i \beta}{6r^{3}r_0^{3}} \right]\\
\rho(r_0)-p_t(r_0) &=& \frac{\lambda_1}{\lambda_2+8\pi} \left[\frac{1}{2r_0^{2}}-\frac{3k}{2r_0^{4}}-\frac{11k\xi_i \beta}{6r_0^{6}} \right].
\end{eqnarray}

\begin{figure}[H]
    \centering
    \includegraphics[scale=.325]{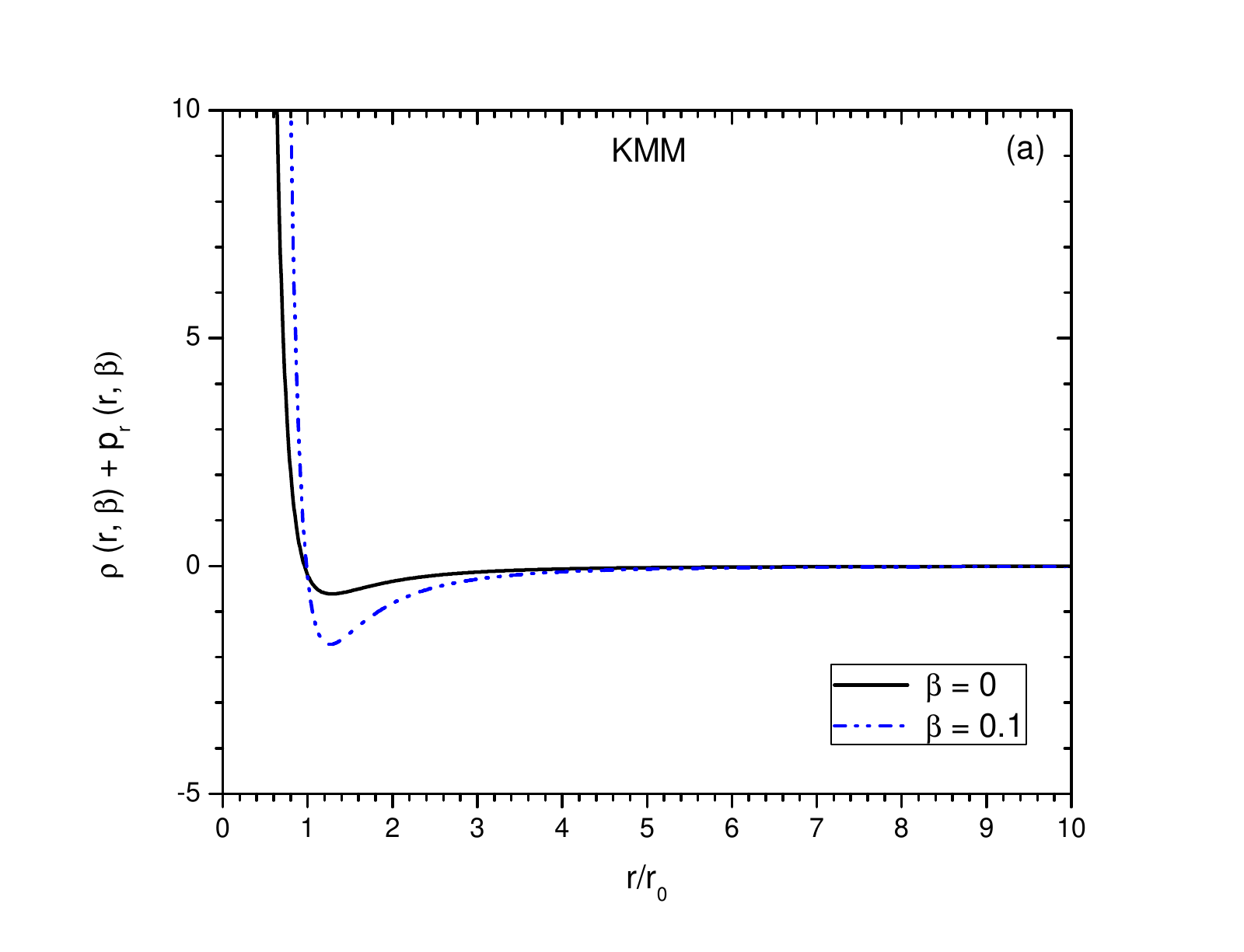}
     \includegraphics[scale=.325]{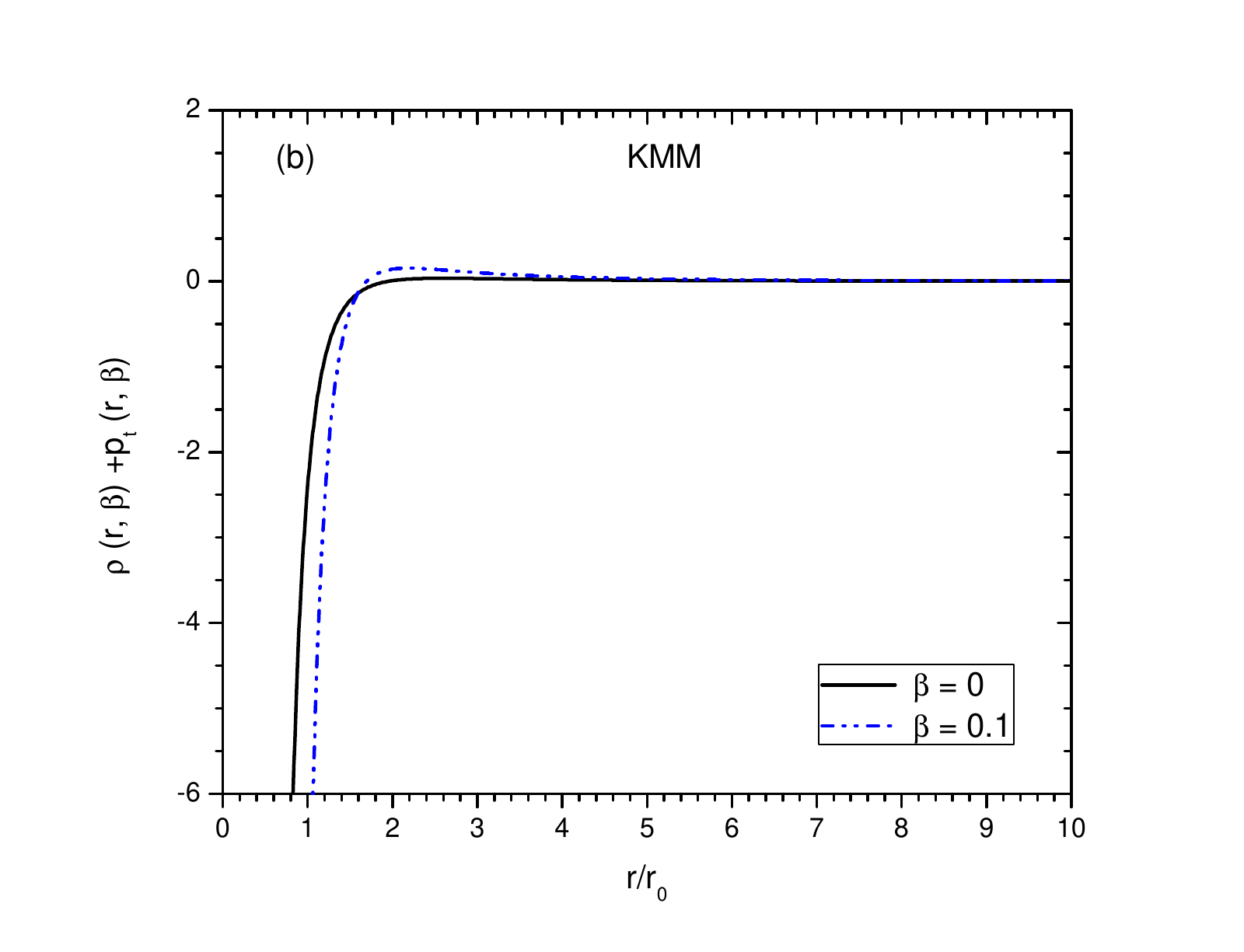}\\
      \includegraphics[scale=.325]{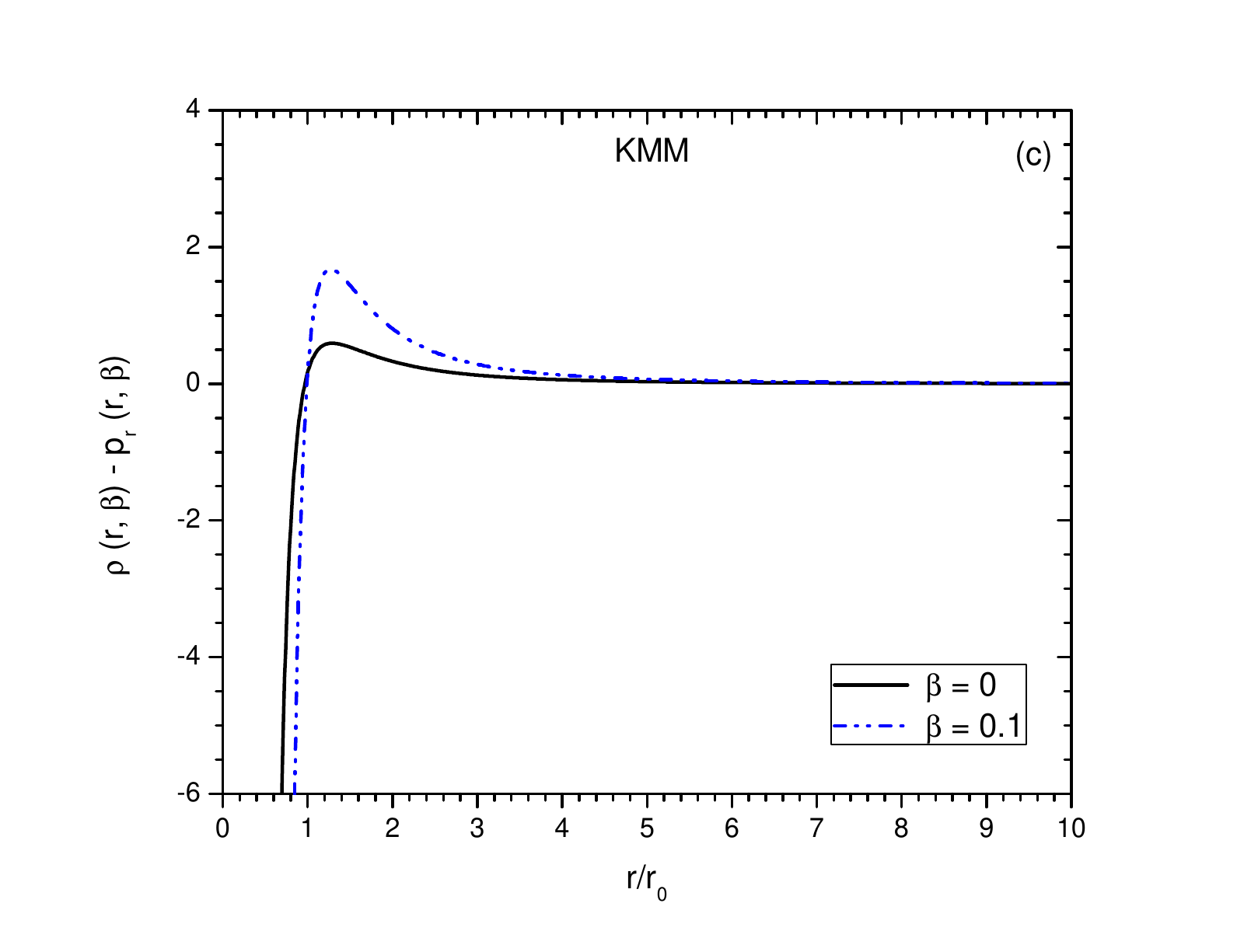}
     \includegraphics[scale=.325]{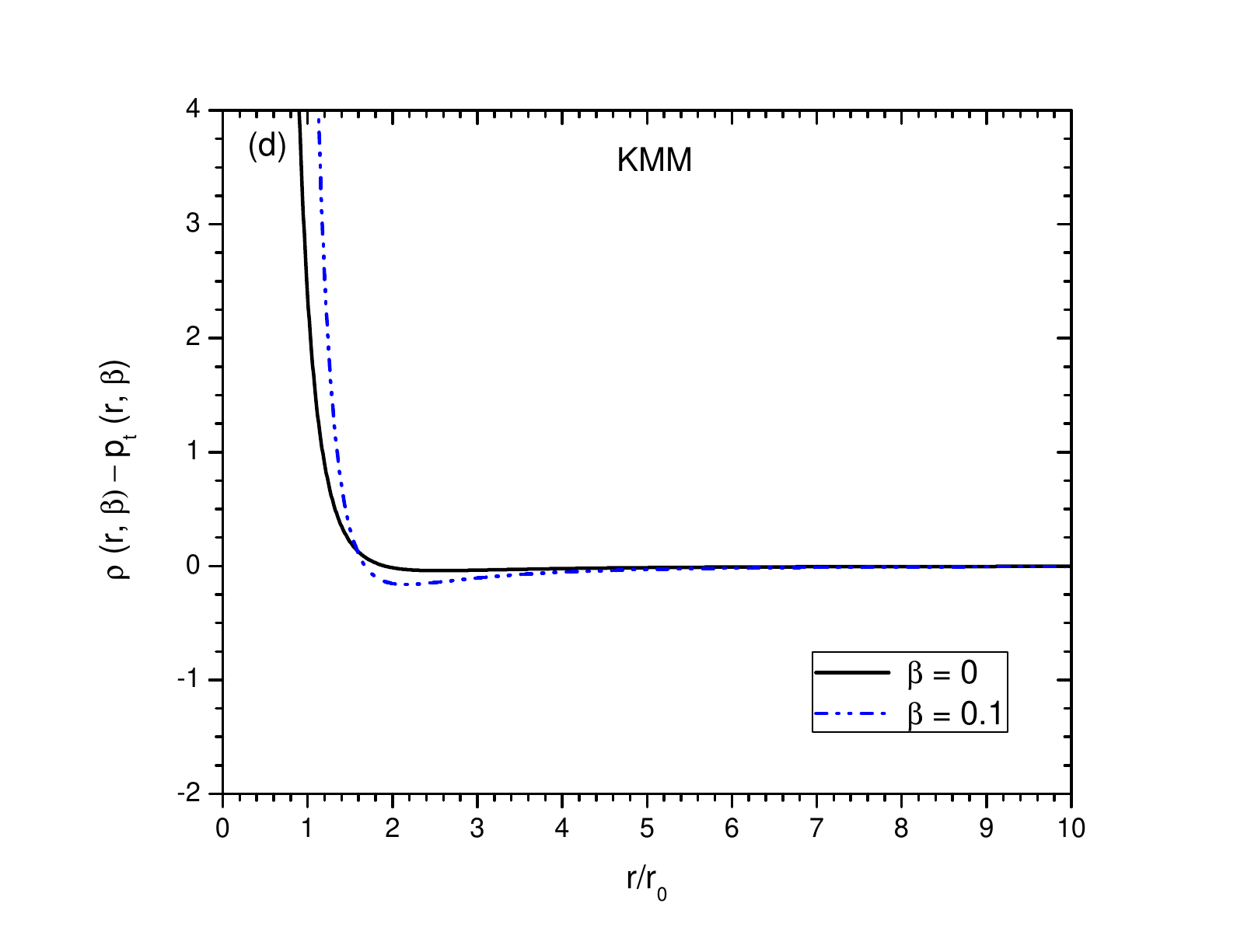}
    \caption{ Different energy conditions and the effect of GUP corrections for KMM constructions. Two different values of $\beta$ namely 0 and 0.1 are considered to observe the effect.}
    \label{fig:6}
\end{figure}

The energy conditions along with the effect of the GUP correction in Casimir wormholes are shown in FIG. \ref{fig:6}. While the NEC1,  i.e, $\rho(r, \beta)+p_r(r, \beta)\geq 0$ is violated beyond the wormhole throat, the NEC2, i.e,  $\rho(r, \beta)+p_t(r, \beta)\geq 0$ is violated within and around the wormhole

%%%%%%%%%%%%%%%%%%%%%%%%%%%%%%%%%%%%%%%%%%%%%%%%%%%%%%%%%%%%%%%%%%%%%%%%%%%%%%%%%%%%%%%%%%%%%%%%%%%%%%%%%%%%%%%%%%%%%%%%%%%%%%%%%%%%%%%%%%%%%%%%%%%%%%%%%%%%%%%
%%%%%%%%%%%%%%%%%%%%%%%%%%%%%%%%%%%%%%%%%%%%%%%%%%%%%%%%%%%%%%%%%%%%%%%%%%%%%%%%%%%%%%%%%%%%%%%%%%%%%%%%%%%%%%%%%%%%%%%%%%%%%%%%%%%%%%%%%%%%%%%%%%%%%%%%%%%%%%%

Traversable wormholes with exotic matter content violate the average null energy condition (ANEC) \cite{Morris88, Hochberg(1998)}. Since quantum effects induce some energy condition violation \cite{Epstein(1965)} , it is pertinent to think of how much ANEC violating matter is present in the spacetime. Visser et al. \cite{Visser03} have proposed a volume integral theorem that quantifies the amount of ANEC violating matter present in the spacetime.\\
Using the integral theorem of Visser et al. we calculate the exotic matter content of the GUP corrected wormholes violating the ANEC as
\begin{equation}
m=\int \left(\rho(r)+p_r(r)\right)dV
\end{equation}
Since $\oint dV=2\int_{r_0}^{\infty} dV=8\pi \int_{r_0}^{\infty} r^{2}dr$, we should evaluate the integral
\begin{eqnarray}
m&=&8\pi \int _{r_0}^{R} \left(\rho(r)+p_r(r)\right)r^{2}dr            \\
&=& 8\pi \int_{r_0}^{R} \frac{\lambda_1}{\lambda_2+8\pi} \left[\frac{r_0}{r^{3}}+\frac{k}{r_0 r^{3}}-\frac{2k}{r^{4}}-\frac{4k\xi_i\beta}{3r^{6}}+\frac{k\xi_i\beta}{3r^{3}r_0^{3}} \right]r^{2}dr      \\
&=&  \frac{8\pi\lambda_1}{\lambda_2+8\pi} \left[\left(r_0+\frac{k}{r_0}+\frac{k\xi_i \beta}{3r_0^{3}}\right)\ln \frac{R}{r_0}+2k\left(\frac{1}{R}-\frac{1}{r_0}\right)+\frac{4}{9}k\xi_i \beta \left(\frac{1}{R^{3}}-\frac{1}{r_0^{3}}\right) \right]         
\end{eqnarray}
One should note that, if $R=r_0$, $m=0$, i.e no exotic matter is required.  However if $R=r_0+\delta$, where $\delta  << 1$ being a small quantity, we have
\begin{eqnarray}
\ln \left(\frac{R}{r_0}\right) &=& \ln \left(1+\frac{\delta}{r_0} \right),\\
\text{and}\\
\frac{1}{R}-\frac{1}{r_0}
&=& \frac{1}{r_0} \left(\frac{1}{1+\frac{\delta}{r_0}}-1 \right),\\
\left(\frac{1}{R^{3}}-\frac{1}{r_0^{3}} \right) &=& \frac{1}{r_0^{3}} \left[\frac{1}{\left(1+\frac{\delta}{r_0}\right)^3}-1 \right],
\end{eqnarray}
so that the exotic matter content of the GUP corrected Casimir wormhole becomes
\begin{eqnarray}
m &=&  \frac{8\pi\lambda_1}{\lambda_2+8\pi}\nonumber\\
  &\times& \left[\left(r_0+\frac{k}{r_0}+\frac{k \xi_i \beta}{3r_0^{3}}\right) \ln \left(1+\frac{\delta}{r_0} \right)+\frac{2k}{r_0} \left(\frac{1}{1+\delta/r_0}-1 \right)\right]\nonumber\\
  &+& \frac{8\pi\lambda_1}{\lambda_2+8\pi}\left[\frac{4}{9} k\xi_i \beta \frac{1}{r_0^{3}} \left(\frac{1}{(1+\delta/r_0)^{3}}-1 \right) \right].
\end{eqnarray}
For $\frac{\delta}{r_0}<<1$,  we get
\begin{equation}
m =-\frac{32\pi}{3} \left(\frac{\lambda_1}{\lambda_2+8 \pi}\right) \left[k\xi_i \beta \frac{\delta}{r_0^{4}} \right].
\end{equation}

\begin{figure}[H]
\centering
       \includegraphics[scale=.325]{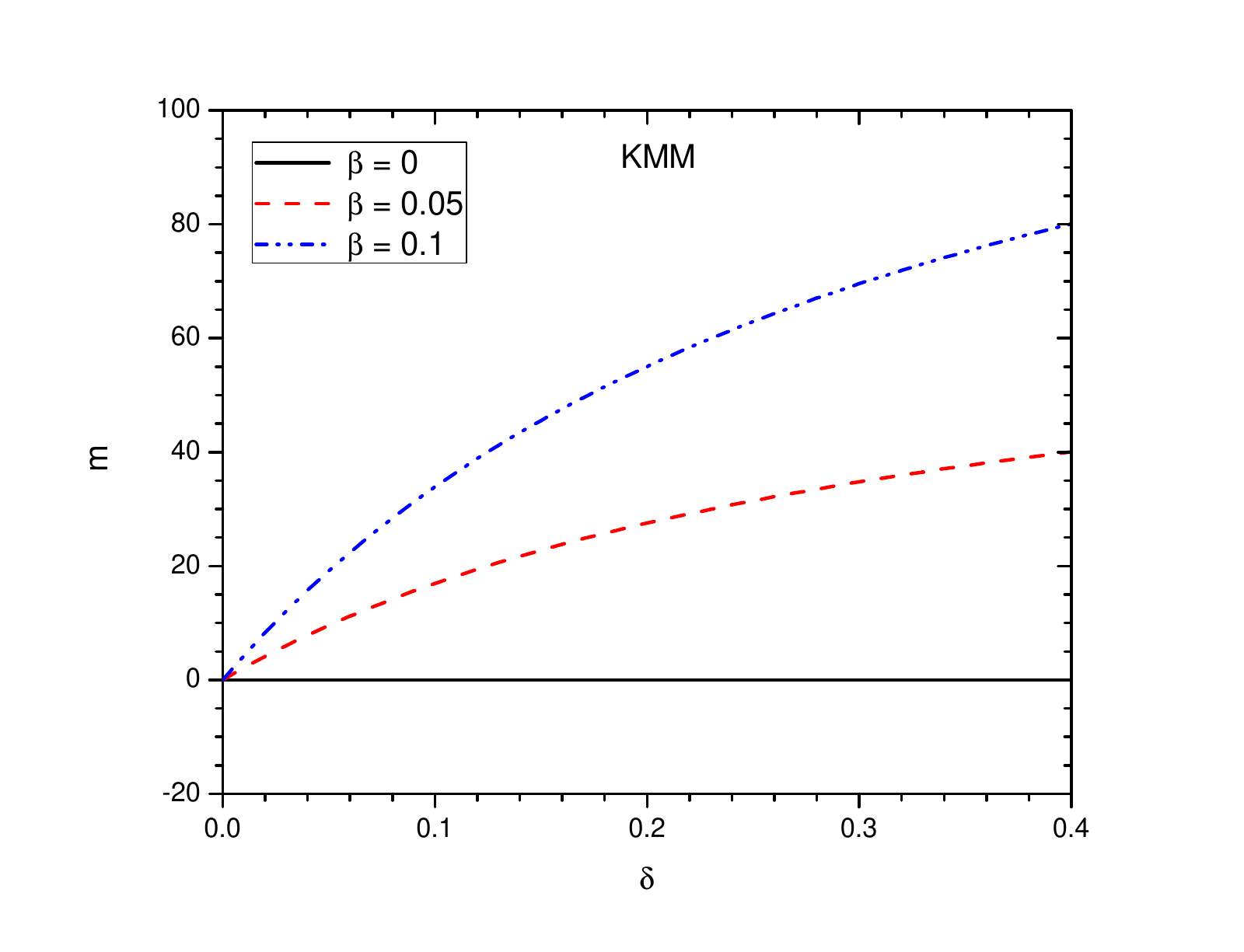}
       \includegraphics[scale=.325]{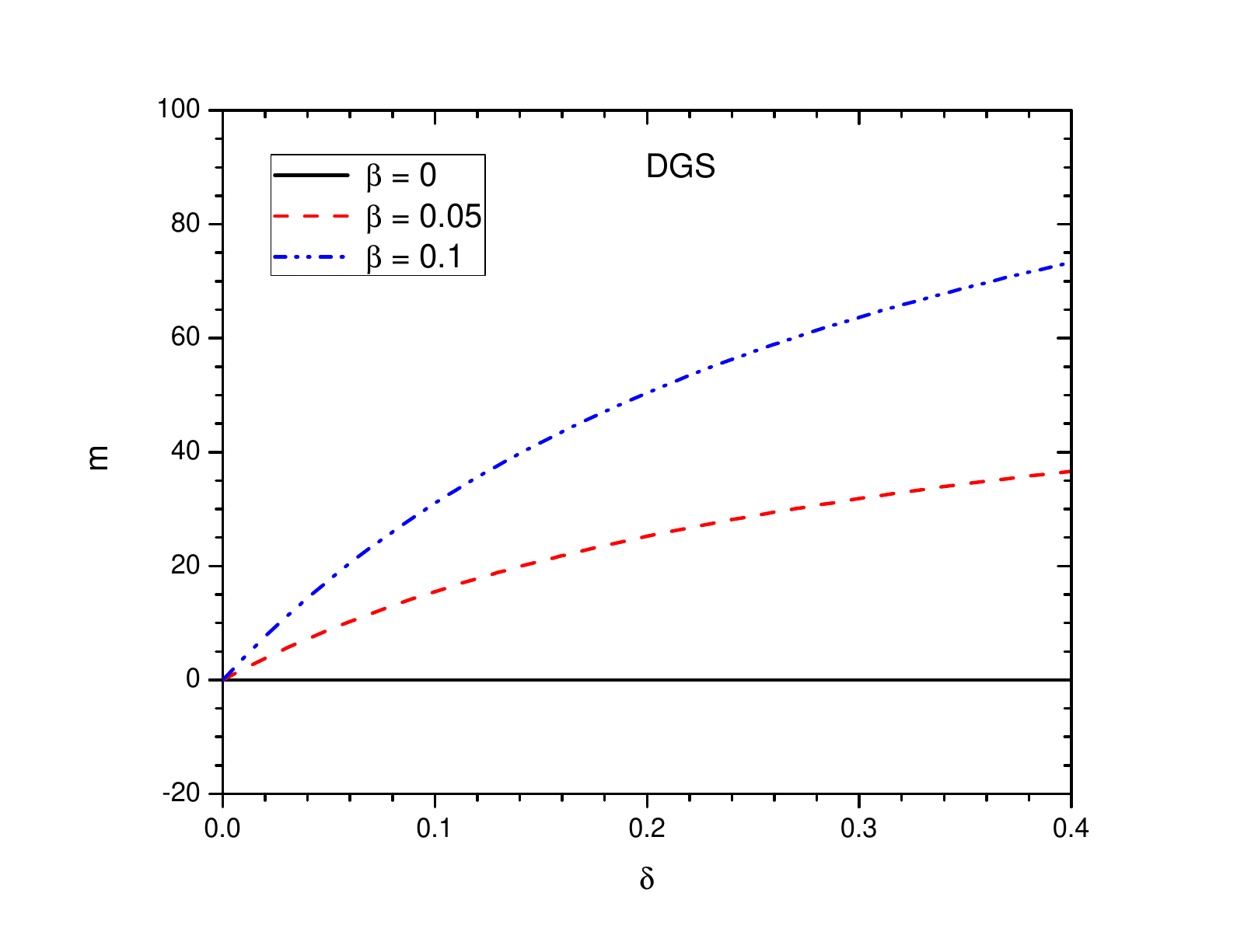}
       \caption{ Exotic matter content of the GUP corrected Casimir wormhole. Left panel for the KMM construction and the right panel for the DGS construction.} \label{fig:7}
\end{figure}

In FIG. \ref{fig:7}, the exotic matter content of the GUP corrected Casimir wormholes is shown for the two constructions of the maximally localized quantum states. 
In other words, a small amount of exotic matter is required to support a traversable wormhole in a region close to the throat. In fact, the total amount of ANEC violating matter can be reduced by considering suitable wormhole geometry. The GUP parameter substantially affects the exotic mass content of the Casimir wormhole. While almost no exotic matter is required for Casimir wormhole with no GUP correction, the exotic matter increases with an increase in $\beta$.

%%%%%%%%%%%%%%%%%%%%%%%%%%%%%%%%%%%%%%%%%%%%%%%%%%%%%%%%%%%%%%%%%%%%%%%%%%%%%%%%%%%%%%%%%%%%%%%%%%%%%%%%%%%%%%%%%%%%%%%%%%%%%%%%%%%%%%%%%%%%%%%%%%%%%%%%%%%%%%%
%%%%%%%%%%%%%%%%%%%%%%%%%%%%%%%%%%%%%%%%%%%%%%%%%%%%%%%%%%%%%%%%%%%%%%%%%%%%%%%%%%%%%%%%%%%%%%%%%%%%%%%%%%%%%%%%%%%%%%%%%%%%%%%%%%%%%%%%%%%%%%%%%%%%%%%%%%%%%%%

\section{Discussion and Conclusion}\label{Sec.V}
In this paper, the role of GUP correction in Casimir wormhole has been presented under the framework of the extended symmetric teleparallel gravity, the $f(Q,T)$ gravity. This modified gravity theory is quite successufl in explaining the late time acceleration phenomenon and other issues in cosmology. Assuming the function $f(Q,T)=\lambda_1 Q+\lambda_2 T$, we have obtained the solution of the modified field equation and discussed the traversable wormhole geometry through the calculation of the shape function. In general the stability 
 issue of traversable wormhole concerning the violation of the null energy condition requires the wormhole matter content to have negative energy density which enables it to open up its mouth so as to make the passage of the physical object to pass through the tunnel. Since such exotic matter with negative energy density  is not possible classically, so a stable as well as traversable wormhole was plausibly unavailalbe. However, quantum mechanical concepts such as the Casimir effect involving the fluctuation of the quantum field near a pair of uncharged, conducting parallel plates provides a hope for realizable matter source with negative energy density. This may well serve as the source for traversable wormholes. In the present work, we explore the possibility of such Casimir energy density to model traversable wormholes within the set up of the $f(Q,T)$ gravity. 

 Another important aspect in quantum mechanics, mostly occurring in supergravity theories and string theories is the concept of the minimal length scale of the order of Planck length leading to the generalization of the usual uncertainty principle. We have applied this GUP correction to the Casimir wormhole and assess its  impact upon the different geometrical and physical properties of the traversable wormhole. In the present work, we restrict ourselves to two different construction techniques of the maximally localized quantum states such as the KMM \cite{Kempf95} and DGS  \cite {Detournay02} constructions. The GUP correction has an exemplified effect on the wormhole shape function outside the wormhole throat in the sense that, with an increase in $\beta$, the shape function is found to increase substantially. The radial and tangential pressures of the Casimir wormhole are also affected by the GUP correction. The pressure anisotropy $\triangle p$ decreases with the increase in $\beta$. It is interesting to note the GUP correction affects the pressure anisotropy only near the wormhole throat while away from the throat, the pressure anisotropy almost vanishes.

The GUP correction to the Casimir wormholes also affects the energy conditions. The null energy conditions are obtained to be violated beyond the wormhole throat and the violation is more strengthened with GUP correction. 

We  have also calculated the exotic matter content of the GUP corrected Casimir wormholes. Almost no exotic mass is required to support the traversable wormhole with no GUP correction, a little amount is required with GUP correction.

In a nutshell, one can notice that, (i) the GUP correction to the Casimir wormholes substantially alters the wormhole geometry and modifies the energy conditions, and (ii) GUP corrected Casimir wormholes require little exotic matter for their stability.

\section*{Acknowledgement}
BM, SKT and SR acknowledge the Inter-University Centre for Astronomy and Astrophysics (IUCAA), Pune, Government of India for providing support through the visiting Associateship program. 
SR also acknowledges the facility availed under ICARD at CCASS, GLA University, Mathura.

\end{document}